\documentstyle[12pt]{article}
\begin{document}
\title{\bf{1/f-noise in the Bak-Sneppen model}}
\author{Frank Daerden, Carlo Vanderzande\\ \\Departement Wiskunde
Natuurkunde Informatica\\Limburgs Universitair Centrum\\3590
Diepenbeek,Belgium}
\maketitle
\begin{abstract}
We calculate time correlation functions in the Bak-Sneppen model (Phys. Rev.
Lett. {\bf 71} 4083 (1993)), a model showing self-organised criticality.
For a random neighbour version of the model, analytical results are
presented, while on a one dimensional lattice we give numerical
results. The power spectrum of these correlation functions shows $1/f$-
behaviour in both cases.
\vspace{1 cm}

PACS-classification: 5.40.+j,64.60.Ak,87.10.+e
\end{abstract}
\newpage
A large diversity of physical systems show $1/f$-noise \cite{Press}. The
power spectra of time correlation functions of such systems show powerlaw
behaviour $f^{-\beta}$ over many orders of magnitude with an exponent
$\beta$ in the range $.6 \sim 1.6$. A possible explanation for the
wide occurence of this phenomenon was put forward in a paper entitled
''Self-organised
criticality: An explanation
of $1/f$ noise`` \cite{BTW1}. In that paper, Bak, Tang and Wiesenfeld
\cite{BTW1} argue that many
open non-linear dynamical systems with large number of degrees of
freedom evolve to a state where they show critical behaviour characterised
by powerlaw correlations both in space and time.
Bak,Tang and Wiesenfeld (BTW) illustrate their ideas using a simple model,
the so called sandpile model \cite{BTW2}. While this model shows
many interesting properties, detailed
investigations \cite{JCF,KK} showed that its power spectrum has $f^{-2}$
behaviour in any finite dimension.
A mean-field calculation of the model did however show the expected
$1/f$-behaviour exactly \cite{TB}.

Following the work of BTW a great variety of models (deterministic and
stochastic, conservative and dissipative, \ldots) have been introduced
which
show the phenomenon of self-organised criticality (SOC). A common feature
of these models is the presence of a separation
of time scales; the system is driven at a very slow rate until one of his
elements reaches a threshold.
This triggers a burst of activity (avalanche) which occurs on a very short
timescale. When the avalanche is over, the system evolves again according to
the slow drive untill a next avalanche is triggered, \ldots. The activity
of the system in this way
consists of a series of independent avalanches. A generic signature of SOC
is the
presence of a powerlaw in the size (or
duration) distribution of the avalanches.
If one increases the external driving rate of the system this powerlaw
disappears. It was however shown by Hwa and
Kardar \cite{HK}, that if one increases the rate at which sand is dropped
in the sandpile model, and one thus obtains
the possibility of interacting avalanches, there appears a region in the
power spectrum where the behaviour is $1/f$.

The BTW-sandpile model is a stochastic and conservative model. Olami,
Feder and Christensen (OFC) \cite{OFC1} introduced
a deterministic and dissipative model, related to spring-block models of
earthquakes, which shows signatures of SOC,
such as the occurence of powerlaw distributions for the sizes of the
avalanches, with an exponent which depends on the
degree of non-conservation in the model. In a subsequent study  \cite{COB}
it was shown that this model shows $1/f$-noise
with an exponent $\beta$ which also depends on the degree of
non-conservation in the model. In a sense then, the OFC-model
fulfills more than the sandpile model the original requirements of the concept
of SOC.

In the present paper we study the question of $1/f$-noise in the
Bak-Sneppen model (BS) \cite{BS}. This model was introduced
to describe the coevolution of species in the Earth's ecology. Indeed the
model shows many qualitative similarities with
data from the real world, but fails on a quantitative level (see e.g.
\cite{FBJS}). In this paper we are only interested
in the BS-model as an interesting physical model and do not discuss its
possible biological relevance.

In the BS-model one has a system of $N$ interacting species, each of which
is represented by a real variable $x_i \in [0,1]$ ($i:1,\ldots,N$)
which is a measure of the fitness of the species.
Initially, all $x_i$ are given a random value, taken from a uniform
distribution on
$[0,1]$. The dynamics of the model is defined as follows.
First one looks for the site $j$ where the fitness takes its lowest value.
One then assigns a new random variable (taken again
from the uniform distribution) $x_j$ to species $j$. At the same time, the
fitness of
$K$ other species is changed randomly. Several
versions of the BS-model can be defined, depending on the way in which
these other species are chosen. In the lattice version
of the model, the species are arranged on a lattice and the $K$ species
are taken as nearest neighbours. A random neighbour
version, in which the $K$ neighbours are chosen at random at each
timestep, was introduced in \cite{FSB}. This version
of the model has the advantage that several of its properties can be
calculated exactly \cite{dDFJW}. In this paper we will
study both this random neighbour version (with $K=1$) and a one
dimensional version of the model in which we only modify
the fitness of the neighbour to the right of the species with lowest
fitness.

Analytical calculations and extensive simulations have shown that the BS-model
evolves to a state in which the probability distribution $p(x)$
that a species has a fitness $x$ becomes a step function, which is zero for
$x$ less then some threshold value $x_c < 1$, and which is
$1/(1-x_c)$ for $x > x_c$. In the random neighbour model it is known that
$x_c=1/(K+1)$ exactly. The exact value of $x_c$ is
not known for any lattice version of the model, but precise numerical
estimates exist, especially in $d=1$, for the
case in which both neighbours are updated \cite{Gr1,PBM}. For the case of
the one-dimensional model in which one neighbour
is updated, we know of no estimate for $x_c$ in the literature.
{}From our numerical results, we estimate $x_c=.710\pm\ 0.005$ for
this case (details of our numerical work are described below).

Once the system has reached the equilibrium state, its dynamics is
characterised by periods (identified with avalanches) in which at least one
of the species
has a fitness less then $x_c$ , separated by periods in which
all species have a fitness above threshold.
The avalanches can be characterised either by their duration or by the
their total activity. Let us denote by $n(t)$ the
number of species which are below threshold as a function of (discrete)
time $t$. The total activity $s$ of an avalanche
lasting from $t=t_-$ to $t=t_+$ (so its total duration is $T=t_+-t_-+1$)
is then given by
\begin{eqnarray}
s\ =\ \sum_{t=t_-}^{t_+} n(t)
\label{1}
\end{eqnarray}
The distributions $P(T)$ of avalanche durations and $P(s)$ of avalanche
sizes follow a powerlaw
\begin{eqnarray}
P(T)\ \sim \ T^{-\tau} \ \ \ ,\ P(s)\ \sim \ s^{-y}
\label{2}
\end{eqnarray}
For the random neighbour model, it is known exactly that $\tau=3/2$
\cite{dDFJW} while for the one dimensional model (2 neighbour
updating) the most accurate numerical estimate is $\tau=1.073 \pm .003$
\cite{Gr1}.
Our simulations of the one-dimensional one neighbour model lead to the
estimate $\tau=1.08   \pm\ .01    $ giving strong evidence that,
as could be expected, both one-dimensional models are in the same
universality class. We don't know of any existing estimates of the exponent
$y$ for the BS-model. We will return to this exponent at the end of the paper.

It is of importance to remark that in the BS-model as described so far
there is no explicit time separation between a fast time scale for
avalanches and a slow time scale for inter-avalanche periods.
Such a separation is however {\it implicitly} present in the definition of the
model since one assumes that one time step in the model is related to a step
in 'geological' time  $t_g=\exp{x_{min}/T}$ (where $x_{min}$ is the lowest
value of $x$ at a given time and $T$ is a measure of mutation rate,
see e.g. \cite{FBJS}). When $1/T\ \gg\ 1$, avalanches occur on time
scales which are short compared to the timescale of the external drive
which is set by the mutation rate.

\vspace{5 mm}
In order to study spectral properties of the BS-model it is necessary to
introduce a dynamical correlation function $G_N(t)$.
In this model a natural definition of a correlation function is
\begin{eqnarray}
G_N(t)\ =\ < n(t_0) n(t_0 + t)>_{t_0} -  <n(t_0)>^2_{t_0}
\label{3}
\end{eqnarray}
where the average is taken over time $t_0$ in the equilibrium state.
According to the dynamical scaling hypothesis \cite{HH} one expects the
Fouriertransform $\hat{G_N}(\omega)$
of a correlation function such as (\ref{3}) to scale as
\begin{eqnarray}
\hat{G_N}(\omega)\ =\ \omega^{-\sigma} H(\omega N^z)
\label{4}
\end{eqnarray}
where $H$ is a scaling function and $z$ the dynamical exponent. Or
equivalently, in real space
\begin{eqnarray}
G_N(t)\ =\ N^{z(\sigma-1)} \tilde{H}(t/N^z)
\label{4a}
\end{eqnarray}
We have calculated $G_N(t)$ analytically for the random neighbour version
($K=1$) of the BS-model and numerically for the
one-dimensional one neighbour version of the model. In both cases we find
the presence of $1/f$-noise.
We now turn to the details of these calculations, and we start with the
analytical results.

In \cite{dDFJW} a master equation approach to the random neighbour model
was introduced. Let $P_n(t)$ be the probability
that at time $t\ , n$ species have a fitness which is below a certain value
$\lambda$. In the end we will be most interested
in the case when $\lambda=x_c$ but for the moment we look at the more
general case.
It is then rather easy to write down a master equation for $P_n(t)$
\begin{eqnarray}
P_n(t+1)= \sum_{m=0}^{N} M_{nm}\ P_m(t)
\label{5}
\end{eqnarray}
where the matrixelements $M_{nm}$ can be written down in terms of $\lambda$
and $N$ \cite{dDFJW}. For $t \to \infty,\ P_n(t)$
evolves to an equilibrium distribution $P_n^*$.
The correlation function $G_N(t)$ can also be written down in terms of the
matrix $M$. One has;
\begin{eqnarray}
G_N(t)\ =\lim_{t_0 \to \infty} \sum_{m=0}^{N}\sum_{k=0}^{N} mk P_m(t_0)
[M^t P(t_0)]_k\ \ -\ \ \left[\lim_{t_0 \to \infty} \sum_{m=0}^N m
P_m(t_0)\right]^2
\label{6}
\end{eqnarray}
This expression in fact allows a (numerically) exact calculation of
$G_N(t)$ in finite systems by simple iteration of the master equation
(\ref{5}).
 We have performed such calculations
for $\lambda=x_c$ for systems with $N$ up to $4000$ and times $t$ up to
$2N$ (results are discussed below).

More interesting is the scaling limit in which $N \to \infty$ and $\lambda
\to x_c$. In that limit it is possible to get a closed
expression for the dynamic correlation function.
It is therefore convenient to rewrite (\ref{6}) as
\begin{eqnarray}
G_N(t)\ =\ \sum_{m=0}^{N}\sum_{k=0}^{N} mk P_m^* Q_{mk}(t)\ -\
\left[\sum_{m=0}^N m P_m^*\right]^2
\label{7}
\end{eqnarray}
where $Q_{mk}(t)$ is the probability that in $t$-timesteps the number of
species with fitness below $\lambda$ changes from $m$ to $k$.
The authors of \cite{dDFJW} assume that in the scaling limit $P_n^*$
becomes a scaling function $f$ of the variable $n/\sqrt{N}$
\begin{eqnarray}
P_n^*\ =\ \frac{1}{\sqrt{N}} f\left( \frac{n}{\sqrt{N}} \right)
\label{8}
\end{eqnarray}
Inserting (\ref{8}) into (\ref{5}) and taking $t \to \infty$, $N \to \infty$
and $\lambda \to x_c$ then gives a differential
equation from which $f$ can be calculated (see eqn. (21) of \cite{dDFJW}).
Using this result we immediately get the second term on the rhs of (\ref{7})
\begin{eqnarray}
[\sum_{m=0}^N m P_m^*]^2\ =\ \frac{N}{2\pi}
\label{9}
\end{eqnarray}
What remains is a calculation of $Q_{mk}(t)$ in the scaling limit.
We therefore assume that this probability scales as
\begin{eqnarray}
Q_{mk}(t)\ =\ \frac{1}{\sqrt{N}}
g\left(\frac{m}{\sqrt{N}},\frac{k}{\sqrt{N}},\frac{t}{N}\right)
\label{10}
\end{eqnarray}
If we insert this assumption in (\ref{5}) and take the scaling limit, we
obtain a differential equation for $g$ (with
$x=k/\sqrt{N}, y=m/\sqrt{N}$ and $\tau=t/N$);
\begin{eqnarray}
\frac{\partial g}{\partial \tau}\ =\ g\ +\ x\frac{\partial g}{\partial x}\
+\ \frac{1}{4}\frac{\partial ^2g}{\partial x^2}
\label{11}
\end{eqnarray}
which has to solved with the initial condition
\begin{eqnarray}
g(x,y,\tau=0)\ =\ \delta(x-y)
\label{12}
\end{eqnarray}
and reflecting boundary conditions in $x=0$.

The solution is
\begin{eqnarray}
g(x,y,\tau)=h(x,y,\tau)+h(x,-y,\tau)
\label{12a}
\end{eqnarray}
where
\setlength{\jot}{6pt}
\begin{eqnarray}
h(x,y,\tau)\ & = & \ \sqrt{\frac{2}{\pi}} \Big(\frac{1}{1-\exp{(-2\tau)}}
\Big)^{1/2} \exp{2y^2} .\nonumber \\
             &   & \exp \bigg\{-\frac{2}{1-\exp{(-2\tau)}}(y^2 + x^2
-2xy\exp{(-\tau)}) \bigg\}
\label{13}
\end{eqnarray}

This result has to be used, together with (\ref{10}), in the first term on
the rhs of (\ref{7}). Taking
the scaling limit and using the expression of $P_m^*$
from \cite{dDFJW} we can rewrite this term as
\begin{eqnarray}
N \frac{2\sqrt{2}}{\sqrt{\pi}} \left[\int_0^{\infty} dx \int_0^{\infty} dy
\ x.y \exp{(-2y^2)} g(x,y,\tau) \right] \nonumber
\end{eqnarray}
Inserting our result for $g(x,y,\tau)$ and performing the integration then
finally gives;
\begin{eqnarray}
G_N(t) & = & N
\Big\{\frac{1}{8\pi}\left(1-\exp{-2\tau}\right)^{3/2}\big[F(1,2,3/2,r_-(\tau))\
+\ F(1,2,3/2,r_+(\tau)) \nonumber \\
       &   & - F(1,2,5/2,r_-(\tau))/3 - F(1,2,5/2,r_+(\tau))/3 \big] -
\frac{1}{2\pi} \Big\}
\label{14}
\end{eqnarray}
where $F(a,b,c,z)$ is the hypergeometric function and where
\begin{eqnarray}
r_\pm(\tau)=\frac{1}{2}\left(1\pm\exp{-\tau}\right) \nonumber
\end{eqnarray}

We thus see that the correlation function has indeed the scaling form
(\ref{4a}) with $z=1$ and $\sigma=2$. In figure 1 we show
our result (\ref{14}) for $G_N(t)/N$ versus $\tau$, together with the
numerical results obtained from direct computation of
(\ref{7}) in finite systems. The agreement is perfect thus lending support
to the
scaling assumptions we made.

To obtain the power spectrum we only have to Fouriertransform (\ref{14}).
Unfortunately, we were not able to obtain an analytical
expression for this transform. The result of a numerical transform using
$^{\copyright}$MATHEMATICA is shown in figure 2. We
show $\hat{G}_N(\omega)/N^2$ versus $\omega N$, which are the natural
scaling variables according to (\ref{4}). The straight
line shown has a slope $-1$. These results then show that over many order
of magnitude
\begin{eqnarray}
\hat{G}_N(\omega)\  \sim\   \frac{N}{\omega}
\label{15}
\end{eqnarray}
so that indeed there is $1/f$-noise in the model.

It is interesting to remark here that the random neighbour versions of
both the BTW-sandpile model \cite{CO} and the BS-model
\cite{FSB} can be related to the critical branching process \cite{Har}.
Within this approximation
both models are thus in the same universality class. Since it is known
that in a mean-field theory the sandpile
model shows $1/f$-noise \cite{TB} it is not so surprising to find the same
results for the BS-model.

We now turn to a discussion of the one-dimensional one neighbour version
of the BS-model. Due to long range correlations
which are present between subsequent species that have lowest fitness
\cite{BS} a master equation approach is no longer
possible. So far, the only approach known for these lattice versions of the
BS-model is numerical.
We have therefore performed extensive numerical calculations of the model
on one dimensional lattices with $N$ up to
$8192$  and for time $t$ up to $2^{32}$. Using these data the values of
$x_c$ and $\tau$ for the one-dimensional
one neighbour model mentionned above were obtained.
Figure 3 shows numerical results for the correlation function $G_N(t)$ for
various system sizes. Surprisingly, for large
system sizes the correlation function seems to become independent of $N$
implying that $z$ becomes $0$. We don't fully
understand this result, but it may be connected with similar behaviour
found for an other exponent ($\eta$) in \cite{PBM}.

Figure 4 shows the power spectrum of the correlation function for the
system with $N=8192$. As can be seen the behaviour
is of the form $\omega^{-\beta}$ over many orders of magnitude. We
estimate $\beta=.97 \pm \ .05$. Thus contrary to
the sandpile model, the BS-model has $1/f$ behaviour also in a lattice
version of the model. The exponent $\beta$ is
furthermore remarkably close to its mean-field value.

We conclude by interpreting our results in the light of a general theory
for $1/f$-noise for systems with self-organised criticality, put forward
in \cite{JCF,KK}. We therefore have to introduce first one more exponent,
denoted $\mu$ which relates the (average) duration $<T>_s$ of an avalanche to
its size $s$ as:
\begin{equation}
<T>_s\  \sim\  s^{\mu}
\label{16}
\end{equation}
In \cite{KK} it is shown on quite general grounds that when $2\mu + \tau>3$
a model shows $1/f$-noise with
\begin{equation}
\beta\ =\ \frac{3 - \tau}{\mu}
\label{17}
\end{equation}

The value of $\mu$ for the one-dimensional BS model can be obtained from
\cite{Gr1,PBM} as follows.
Let $<n(t_a)>$ denote the average number of species below threshold
a time $t_a$ after the start of an avalanche.
In \cite{Gr1,PBM} it is shown (numerically) that  $<n(t_a)>$ grows slower
then any power (and even becomes
a constant according to \cite{PBM}). Therefore, for long living avalanches, it
follows from (\ref{1}) that $<T>_s \ \sim\ s$, or $\mu=1$ for the
one-dimensional model.
Turning to the random neighbour BS-model, we don't know how to calculate
$\mu$ exactly from the master quation approach to this model. From the
precise form of the transition probability matrix $M$ as derived in
\cite{dDFJW},
it is however clear that for $\lambda=x_c, n(t)$ performs a random walk, from
which it can be concluded that for long times after the start of
an avalanche $<n(t_a)>\  \sim\  t^{1/2}$, so that using (\ref{1}) we obtain
$\mu=2/3$.
We have indeed verfied this $\mu$-value in simulations of the random neighbour
BS model.
(This is an appropriate place to remark that these estimates of $\mu$ allow
us to determine the exponent $y$ (see (\ref{2}) from the obvious relation
$y=[(\tau-1)/\mu] + 1$ so that for the one dimensional BS-model we obtain
$y=\tau$ while for the random neighbour model $y=7/4$.)

Using these values we find  for the random neighbour version
of the BS-model $2\mu + \tau=17/6 < 3$ so that the results of
\cite{KK} predict $\beta=2$, or absence of $1/f$-noise. For the one dimensional
model the results of this reference do predict $1/f$-noise but with an
exponent $\beta \approx 1.92$. Both predictions are in contradiction
with the numerical results which we presented here.
A possible, heuristic, explanation for this is the following. In the
arguments of
\cite{KK} time correlations {\it within one avalanche} are considered.
This is the natural thing to do, since in a generic SOC-model the avalanches
occur instantly on the long time scale, and therefore correlations including
the inter-avalanche period would be trivially zero.
As argued above this explicit time separation is not present in the BS-model.
Our calculation of the time correlation function therefore include
periods in which their is no activity. These lead to a more rapid decorrelation
as compared to the case where one only studies intra-avalanche correlations.
This more rapid decorrelation in turn leads to a decrease in the power of the
low frequency components of the power spectrum and therefore to a
$\beta$-exponent
which is lower then that following from simple application of the results
in \cite{JCF,KK}.
In a sense then, the BS-model is somewhat akin to the sandpile model with
a finite driving rate as studied in \cite{HK}, but without the possibility
of interacting avalanches.
This mechanism seems to lead to a model which shows both powerlaws in
space and time correlations without destroying the powerlaws in the
size (and/or time) distribution of avalanches.
It is an interesting project to investigate the generality of this scheme.

{\bf Acknowledgement} We thank the Program on Inter-University Attraction
Poles, Prime Minister's Office, Belgian Government for financial support.

\newpage
{\Large\bf Figure Captions}

\underline{\bf Fig.1} The exact correlation function of the random
neighbour Bak-Sneppen model. The figure shows
the exact result (16) together with appropriately scaled finite system
results obtained using (8). The results
are for $N=250(\Diamond),N=500(\times),N=1000(\Box),N=2000(+)$ and
$N=4000(\triangle)$.

\underline{\bf Fig.2} Numerical Fouriertransform (open circles) of the
exact correlation function (12). The straight line represents
a best fit through the linear part of the data and has a slope of -1.

\underline{\bf Fig.3} Numerical results for the correlation function of the
one-dimensional one neighbour Bak-Sneppen model.
The different curves represent results for (bottom to top) $N=128,256,1024$
and $4096$ respectively. The upper two curves
almost completely coincide.

\underline{\bf Fig.4} Fouriertransform (crosses) of the numerically
calculated correlation function of the one-dimensional
one neighbour Bak-Sneppen model. The results are for a system of $N=8192$
species. The straight line represents a best
fit through the linear part of the data and has a slope of -.972.

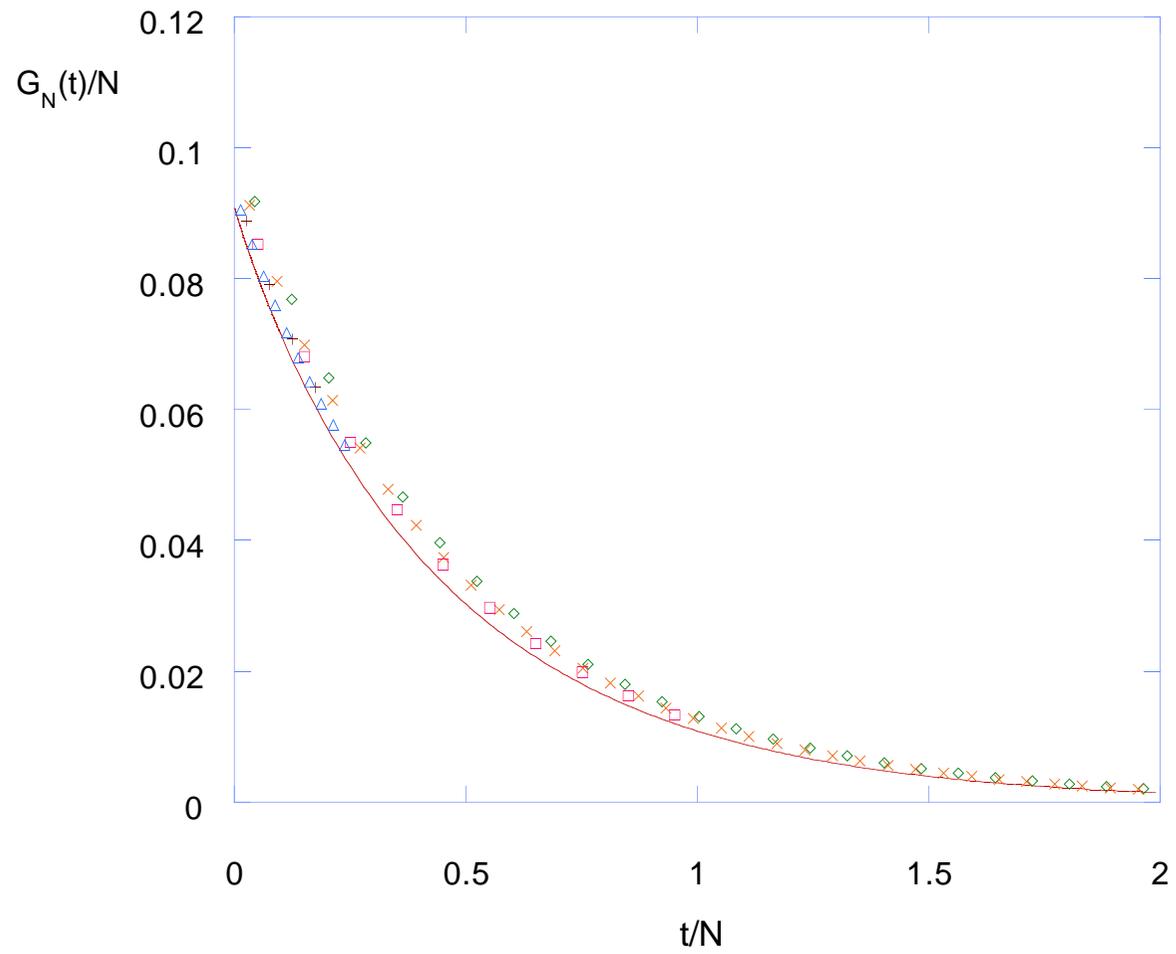
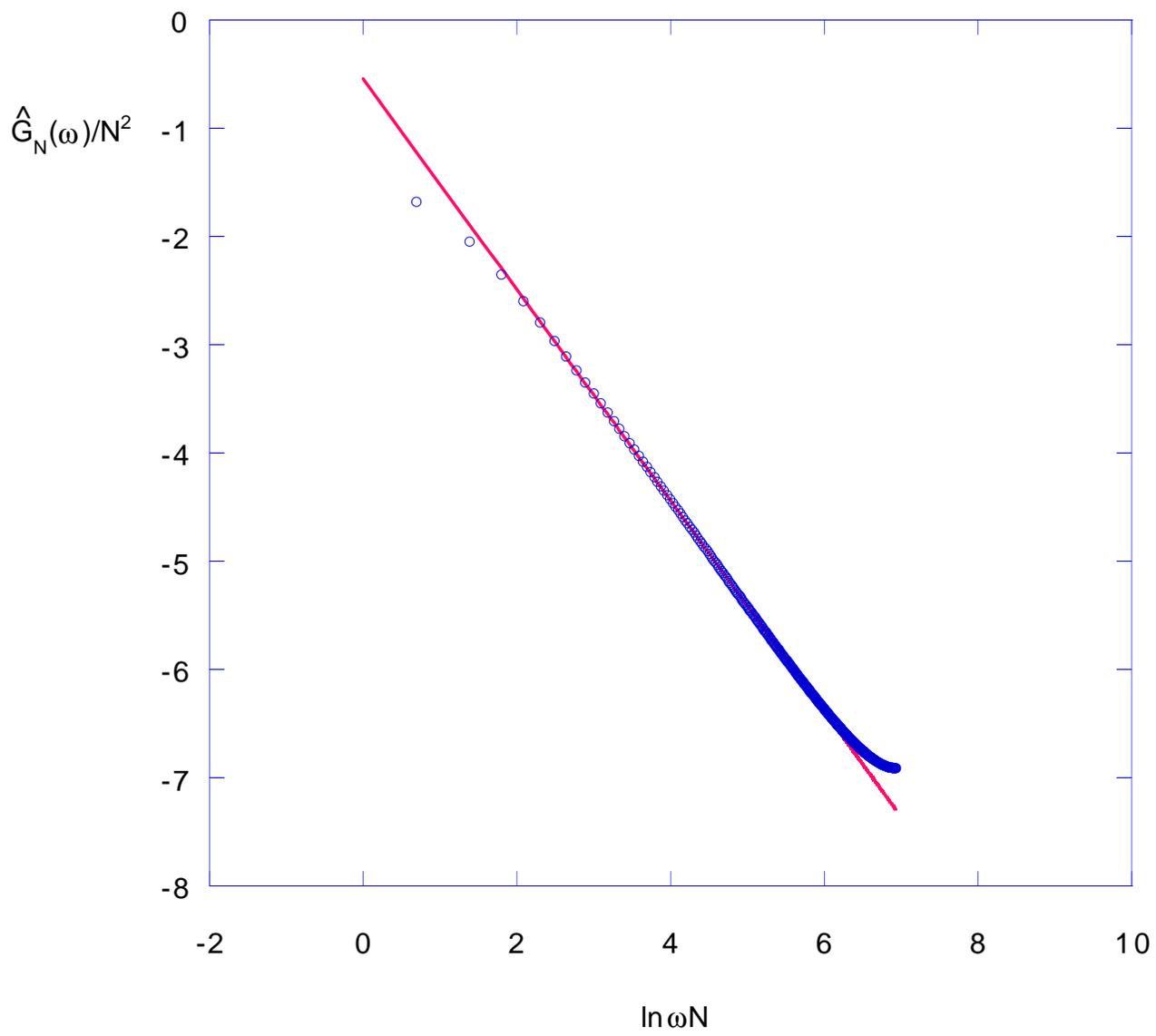
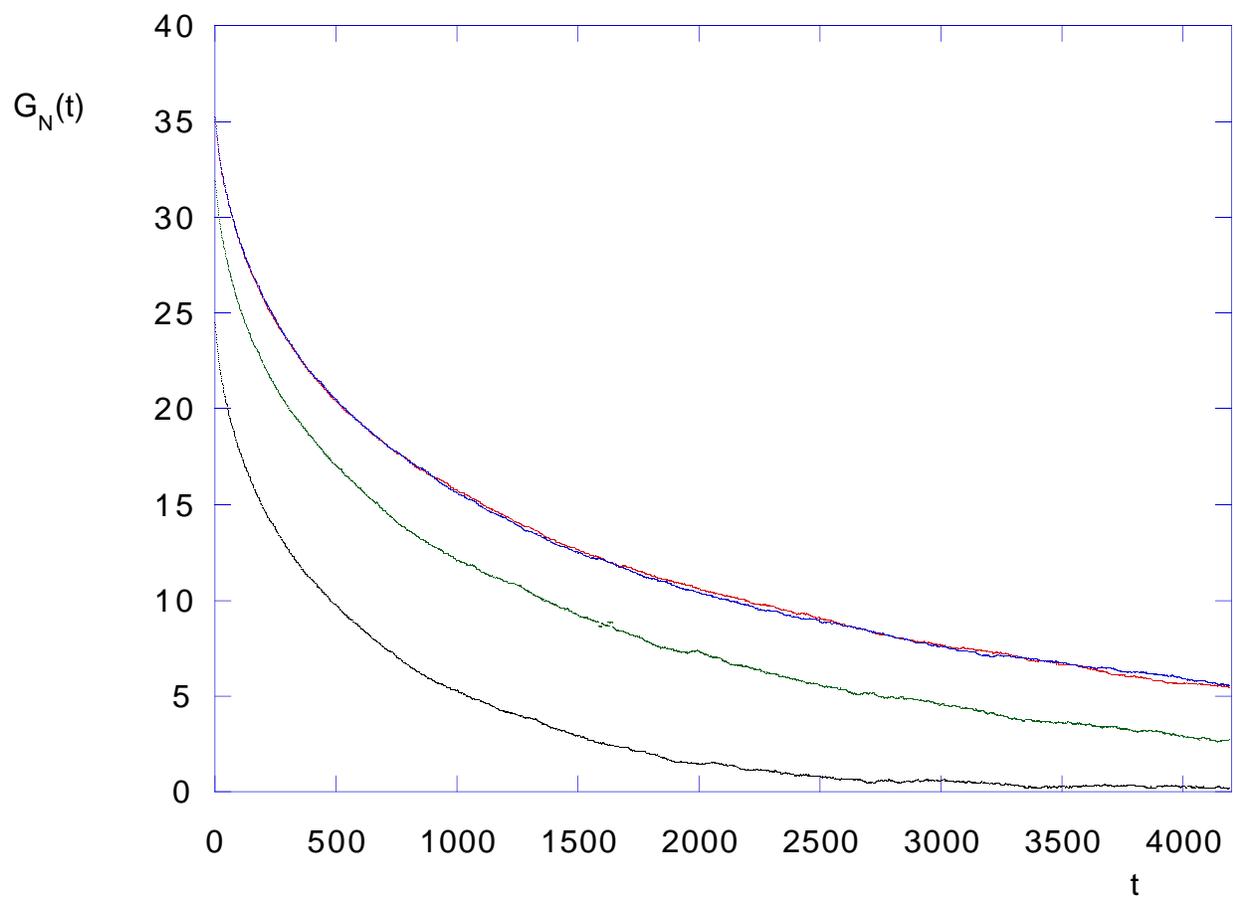
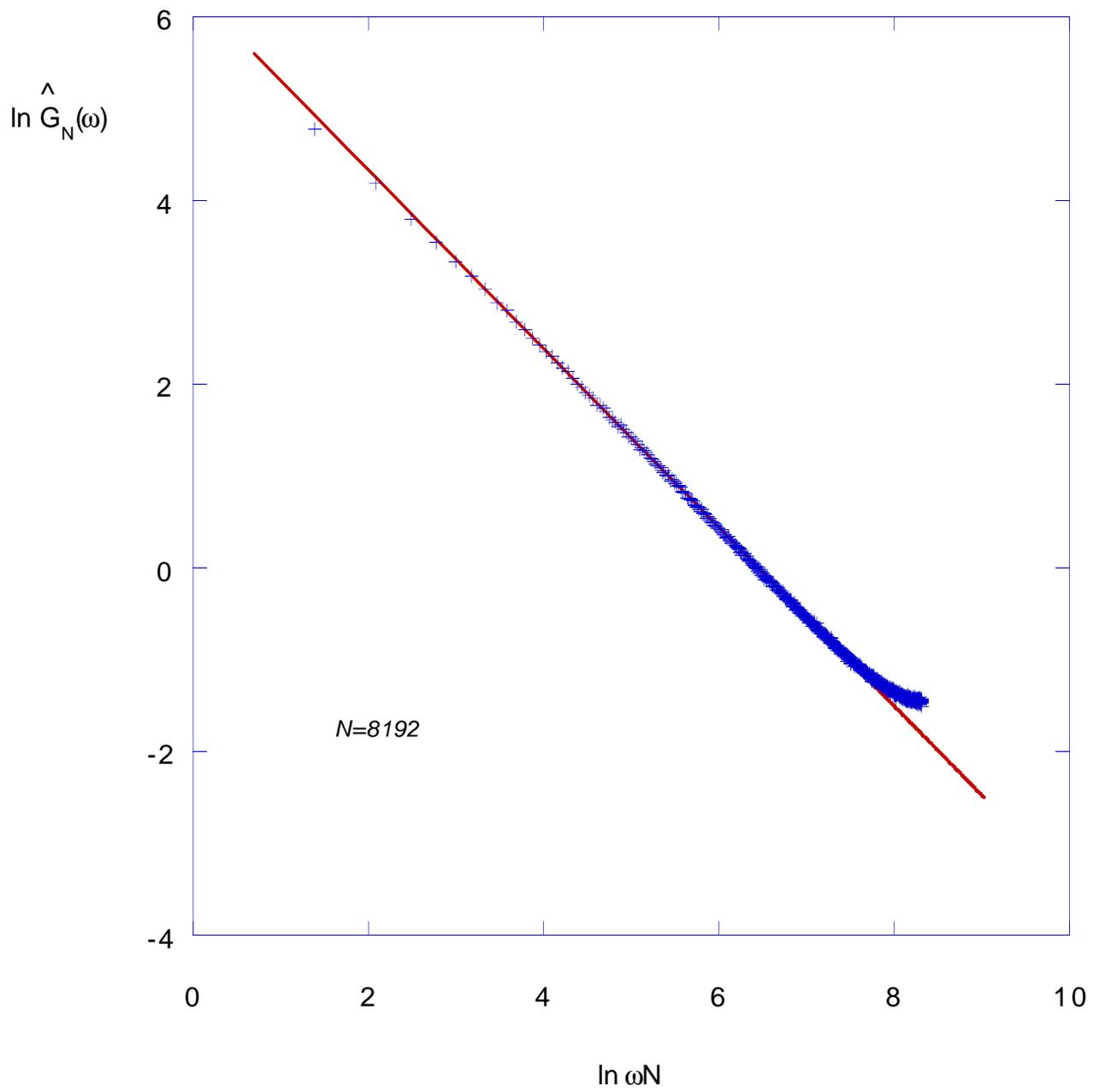
\end{document}